\documentclass[prl,twocolumn]{revtex4-1}

\usepackage{graphicx}
\usepackage{amsmath}
\usepackage[dvips]{color}

\begin{document}

\title{Vortex Dynamics in Anisotropic Traps}

\author{S.~McEndoo and Th.~Busch}

\date{\today}

\begin{abstract}
  We investigate the dynamics of linear vortex lattices in anisotropic
  traps in two-dimensions and show that the interplay between the
  rotation and the anisotropy leads to a rich but highly regular
  dynamics.
\end{abstract}

\pacs{03.75.Kk, 67.85.De}

\affiliation{Physics Department, University College Cork, Cork, Ireland}

\maketitle

\section{Introduction}

Since their first observation in solid state superconductors
\cite{Essman:67} and liquid $^4$He \cite{Yarmchuk:79} arrays of
quantized vortices have been one of the key features of systems
exhibiting frictionless flow. More recently the observation of vortex
lattices in gaseous Bose-Einstein condensates
\cite{Dalibard:00,AboShaeer:01} has led to an re-newed interest in
these systems, as ultracold atomic gases are highly configurable
systems and allow for investigation of the lattices under a wide range of
external parameters. Since then a large number of theoretical and
experimental studies have been undertaken, which, for example, have
examined the equilibrium properties of vortex lattices in different
external potentials \cite{Reijnders:04,Aftalion:04} or the dynamical
properties of vortices lattices when disturbed
\cite{Coddington:03}. At the same time new ideas for applications in
quantum information \cite{Kapale:05,Thanvanthri:08} and precision
measurement \cite{Thanvanthri:09} have been developed.

The geometry of a vortex lattice is determined by the condition to
minimize the energy of the overall system under the provision of fixed
angular momentum. It therefore depends on the interaction energy
between the velocity fields of the vortices, which is a logarithmic
function of the distance between, and the underlying external
potential. For Bose-Einstein condensates carrying large angular
momentum in isotropic harmonic traps this leads to the formation of a
triangular Abrikosov lattices of vortices with winding number
one~\cite{AboShaeer:01,Anglin:02}. For trap geometries that are not
simply connected, such as a toroidal shapes, the system can also
responds by forming persistent currents with higher winding numbers
\cite{Ryu:07}. Other trap geometries and parameters, such as
anisotropic traps \cite{Oktel:04,Sinha:05,McEndoo:09} or optical
lattices \cite{Reijnders:04}, no longer support an Abrikosov lattice,
but instead form zigzag or linear, and square lattices respectively.

While vortices and vortex lattices are interesting to study on their
own merits \cite{Fetter:01,Fetter:09}, it has been suggested that they
can be used in quantum information applications through encoding of
information into the winding number
\cite{Kapale:05,Thanvanthri:08}. These numbers are so-called
geometrical quantum numbers, which have the advantage of not suffering
from any energetic instabilities and which are stable as long as the
vortex cores are part of the system. Transferring the atoms into
superposition states of these quantum numbers can be done either by
optical methods \cite{Kapale:05}, or by making use of anisotropic
external potentials \cite{Perez-Garcia:07}. However, due to the very
dense energy spectrum of atomic superfluid systems it is a very
difficult task to separate out exactly two flow states
\cite{Hallwood:07,Nunnenkamp:08}. Since the vortex-vortex interaction
is dependent on both magnitude and direction of rotation, however, one
can in principle imagine a setup for a superfluid quantum computer
analogous to the linear ion-trap model of Cirac and Zoller
\cite{Cirac:95}. In order to facilitate control and addressing of the
individual vortex qubits it therefore would be preferable to reduce
the number of nearest neighbours from six in the Abrikosov lattice,
to, say, two in a linear crystal. It is known by today that two
systems with which this can be achieved are waveguides \cite{Sinha:05,
  Sanchez-Lotero:05} and anisotropic traps \cite{Oktel:04,McEndoo:09}.

Here we consider the latter systems at medium rotation and with a
small numbers of vortices. For currently typical experimental
non-linearities it was recently found that even relatively small
anisotropies, $\lambda \approx 2$, lead to the vortices forming a
linear crystals with equal spacing between them
\cite{McEndoo:09,LoGullo:10}. However, the inhomogeneity of the
background gas leads to a finite velocity of the vortex lattice within
the cloud \cite{Feynman:55}, which in the limit of large vorticity
cancels out the external rotation \cite{AboShaeer:01}. For systems of
medium vorticity in an anisotropic trap it is obvious that the linear
shape of the crystal cannot be maintained during these dynamics and we
will show here that this results in excitations within the background
cloud which can be described by a standard hydrodynamics approach
\cite{Cozzini:03,Lobo:05}. More surprisingly, we also find that the
excitations are oscillatory and do not destroy the ordering of the
vortex lattice, even in the situation when the external rotation is
switched off.

In the following we will first briefly introduce our model and then
examine the dynamics of the linear vortex lattices as a function of
anisotropy by integrating the underlying Gross-Pitaevskii equation for
time scales of several hundred milliseconds. We determine the
characteristic excitation frequencies of the vortex lattice and
compare them to the known frequencies for anisotropic
condensates. Finally, we remark on the stability of the vortices
themselves.

\section{Anisotropic Trapping Potentials}

To simulate the dynamics of the vortices we consider a two-dimensional
model of Bose condensed gas of particles of mass $m$ in a harmonic
trap, $V_\text{HO} =\frac{1}{2} m(\omega_x^2x^2 +\omega_y^2 y^2)$,
where the anisotropy is determined by the aspect ratio,
$\lambda=\omega_y/\omega_x$. The trap is rotating around the $z$-axis
and we assume that a strong potential suppresses all relevant dynamics
in this direction. For small, but non-negligible anisotropies
($\lambda \gtrsim 2$ for typical experimental systems), the ground
state for medium rotation frequencies is given by a single line of
vortices along the weak axis of the trap \cite{McEndoo:09}. In the
case of an odd number of vortices, one vortex will be located at the
centre of the trap and the others will be symmetrically arranged
around it. For an even number the same symmetry exists, however
without the central vortex. Furthermore the vortex spacing remains
roughly even along the soft axis due to the slow variation of the
condensate density.

For large rotational frequencies the condensate's efforts to
distribute the vorticity as uniformly as possible lead to a
self-consistent solid body rotation which cancels out the overall
rotations of the lattice in the co-rotating
frame. \cite{Feynman:55}. In the region of medium vorticity, however,
in which only a few vortices are present, the concept of a diffused
vorticity cannot be applied and the vortex lattice will try to carry
out rotation in the co-moving frame. It is not surprising that in
anisotropic traps this will lead to a dynamics that will have a
feedback onto the density distribution of the atomic cloud.

In the following we will consider these dynamics as a function of the
trap anisotropy for the case of $^{87}$Rb and currently realistic
trapping frequencies. All our simulations are done by
numerically integrating the Gross-Pitaevskii equation using a
FFT/split-operator method.

\section{Dynamics as function of anisotropy}

The Hamiltonian for a Bose-Einstein condensate in an anisotropic trap
and under rotation is given by
\begin{align}
  \label{eq:hamiltonian}
  H = -\frac{\hbar^2}{2m} \nabla^2 + V_\text{HO} -\Omega L_z + \frac{gN}{2} |\psi |^2\;,
\end{align}
where $N$ is the number of atoms and $\Omega$ is the frequency with
which the trap rotates around the $z$-axis. As usual, for stability
reasons $\Omega$ is restricted to values smaller than the trapping
frequency of the soft axis of the condensate. The two dimensional
interaction strength is given by $g=(4\pi\hbar^2a_{sc}/m)
(m\omega_z/2\pi\hbar)^{1/2}$, with $a_{sc}=4.67$ nm being the three
dimensional scattering length for $^{87}$Rb in the $|2,1\rangle$
state.

We are interested in studying the behaviour of a ground state lattice
as a function of time.  While the behaviour of the condensate as a
whole can be studied using a hydrodynamics approach in the
Thomas-Fermi limit \cite{Cozzini:03,Lobo:05}, the dynamics of the
vortices themselves require an approach that is able to resolve the
dynamics on the length scale of the healing length, such as the
Gross-Pitaevskii equation. To have a clean initial state we therefore
first find the ground state of the Hamiltonian~\eqref{eq:hamiltonian}
by a numerical relaxation method and then integrate this state in time
while maintaining the rotation.

\begin{figure}[tb]
  \includegraphics[width=\linewidth,clip=true]{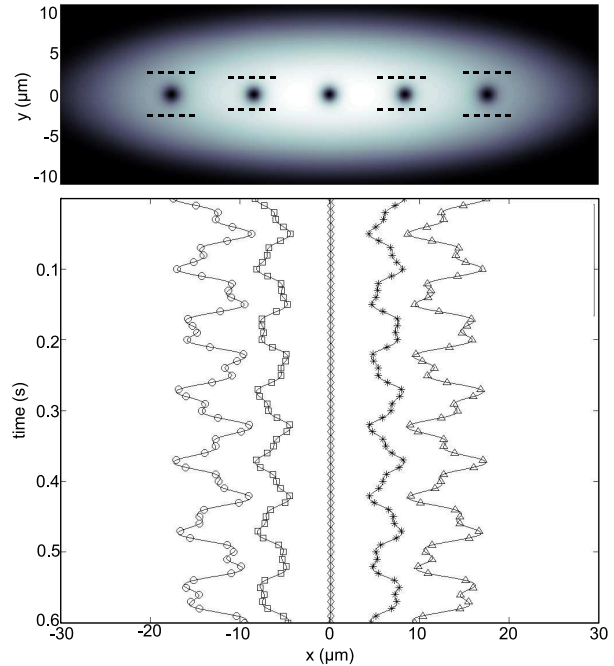}
  \caption{The ground state density for a condensate of $N=10^5$ Rb
    atoms in a trap with $\lambda = 2.5$ and $\omega_x = 10\times 2\pi
    \;\text{Hz}$ [upper panel] and the oscillatory trajectories of the
    vortices along the $x$-axis as a function of time [lower
    panel]. The centre vortex can be seen to remain stationary, while
    the other four vortices periodically oscillate away from their
    initial position. The dashed lines in the density plot indicate the
    range of movement in the $y$-direction for each of the vortices
    (see Fig.~\ref{fig:lambda2p5y}). }
\label{fig:lambda2p5x}
\end{figure}

\begin{figure}[tb]
  \includegraphics[width=\linewidth,clip=true]{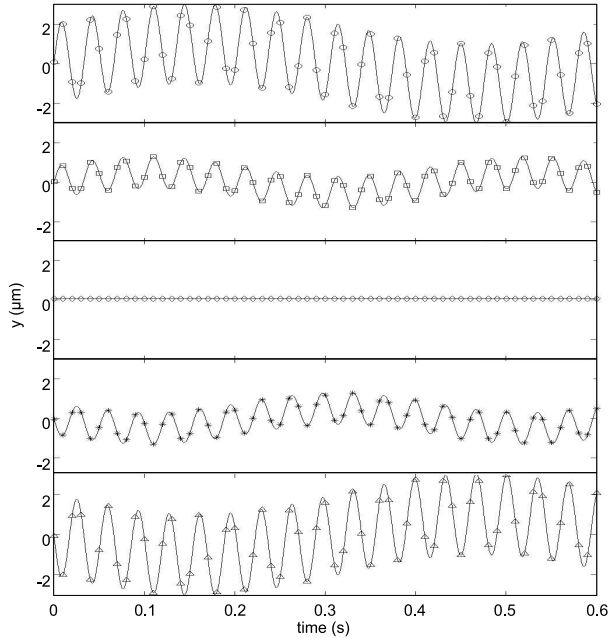}
  \caption{The position of the vortices in the $y$-direction
    corresponding to the situation shown in Fig.~\ref{fig:lambda2p5x}.
    The top row shows the leftmost vortex and the bottom row the
    rightmost one.
  }
\label{fig:lambda2p5y}
\end{figure}

The results of the numerical integration for a crystal of five
vortices in a trap of aspect ratio $\lambda=2.5$ are shown in
Fig.~\ref{fig:lambda2p5x}. The upper half shows the starting
configuration and the lower half displays the positions of the
individual vortices as a function of time along the soft direction of
the trap. The detailed displacement in the tight direction of the trap
is shown in Fig.~\ref{fig:lambda2p5y} and its range is indicated by
the dashed lines in Fig.~\ref{fig:lambda2p5x}.

\begin{figure}[tb]
  \includegraphics[width=\linewidth,clip=true]{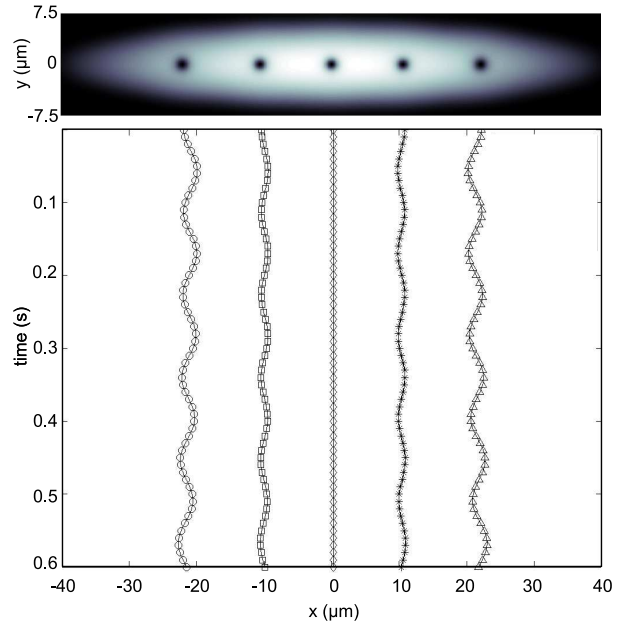}
  \caption{The ground state density for $\lambda = 3.5$ [upper panel],
    and the position of the vortices along the $x$-axis [lower
    panel]. All system parameters are the same as in the $\lambda =
    2.5$ case and all observations made apply here as well with
    reduced amplitude. In the $y$-direction the movement is on the
    order of $1\times 10^{-7}$ m (not shown). }
\label{fig:lambda3p5x}
\end{figure}

There are four immediate observations one can make from looking at
these results: (i) the trajectories in both directions are symmetric
for the pairs of vortices that are at equal distance from the centre
of the trap. The central vortex remains stationary at all times; (ii)
while the oscillation amplitudes for the two inner and the two out
vortices are different, they are in phase with each other during the
whole evolution; (iii) the vortex motion is periodic and in particular
the direction of rotation of the lattice reverses regularly and (iv)
the paths of the vortices never cross and the linear ordering is
maintained at all times. Closer examination shows that the
trajectories of the two vortices on the right hand side and the two
vortices on the left hand side follow the exact same pattern and are
only different by a constant factor in their amplitude. This applies
to the motion in the $x$ as well as in the $y$ direction and shows
that the distance between neighbouring pairs of vortices is the same
at any point in time. We have calculated these dynamics for different
aspect ratios and found the behaviour consistently. As an example
Fig.~\ref{fig:lambda3p5x} shows the results for a trap with an aspect
ratio of $\lambda=3.5$, for which the same behaviour, apart from a
smaller amplitude of the oscillations, is found.

To better understand this pattern, we show in Table
\ref{tab:comparison} the dominating oscillation frequencies for the
vortex oscillations in three different traps.  For all anisotropies we
find two dominating frequencies, $\omega_1$ and $\omega_2$, and in the
soft direction the lower frequency has a larger amplitude whereas in
the tight direction the opposite is true. For increasing anisotropy
the lower frequency decreases, whereas the higher frequency increases.
This can clearly not be explained by straightforward oscillations in
the harmonic potential, but indicates that the non-linearity of the
sample plays a significant role. At the same time, it is also clear
that the oscillation does not originate from a simply stationary
vortex crystal inside an oscillating background cloud, as the cloud
would be stationary in the co-rotating frame. However, we will show in
the following that both factors are playing an important role.

\begin{table}[b]
\begin{center}
\begin{tabular}{|c|c|c|c|c|c|c|}
\hline
$\lambda$ &$\omega_x$&$\omega_y$&$\omega_1$&$\omega_2$&$\omega_{h1}$&$\omega_{h2}$\\
\hline
2.5 & 62.8 Hz & 157.1 Hz & 69.0 Hz & 187.9 Hz & 69 Hz & 182 Hz \\
3.0 & 62.8 Hz & 188.5 Hz & 60.0 Hz & 210.0 Hz & 67 Hz & 210 Hz \\
3.5 & 62.8 Hz & 219.9 Hz & 53.7 Hz & 254.4 Hz & 62 Hz & 239 Hz \\
\hline
\end{tabular}
\caption{\label{tab:comparison} Frequencies dominating the oscillation
  of the vortex lattice from the numerical integration of the
  Gross-Pitaevskii equation ($\omega_1$ and $\omega_2$) and
  oscillation frequencies from the hydrodynamic model for the
  background cloud ($\omega_{h1}$ and $\omega_{h2}$) for different
  values of the anisotropy. For comparison the trapping frequencies
  are shown as well.}
\end{center}
\vspace{-0.6cm}
\end{table}

Let us describe the dynamics of the background cloud using a
hydrodynamic formalism and assume that the excitation of it are
triggered and driven by the motion of the vortices. In fact, we assume
that the rotation of the vortices leads to an infinitesimal rotation
of the whole cloud and solve the classical hydrodynamical equations
from there. This scenario was recently investigated by Lobo and Castin
in order to find the scissors mode of a rotating condensate with
vortices \cite{Lobo:05}.

The density profile, $\rho({\bf r})$, and the velocity field, ${\bf
  v}({\bf r})$, of a condensate in a rotating anisotropic trap in the
Thomas-Fermi limit in the co-rotating frame can be described by the
equations of classical hydrodynamics as
\begin{align}
  (\partial_t+\bf{v}\cdot\nabla){\bf v}=&
                -\frac{1}{m}\nabla(U+\rho g)
                -2{\bf \Omega}\times{\bf v}\;,\nonumber\\
  &-{\bf \Omega}\times({\bf \Omega}\times{\bf r})\\
  \partial_t\rho+\nabla\cdot\rho{\bf v}=&0\;.
\end{align}
Because we have chosen the ground state with a stationary linear
vortex lattice as the initial condition for the cloud, the condensate
is characterised by ${\bf v}\approx 0$ for the velocity field and by
the Thomas-Fermi solution for the density
\begin{equation}
  \rho_0(x,y)=\frac{m}{2g}\left[2\mu-(\omega_x^2-\Omega^2)x^2
                                    -(\omega_y^2-\Omega^2)y^2)\right]\;.
\end{equation}
We assume that the rotation of the vortex lattice, and therefore the
change in the associated flow, leads to a small, initial disturbance
that corresponds to a rotation of the whole system with respect to the
trap. It can be described by a small angle, $\theta$, from which the
initial conditions for the linear perturbations to the above equations
are given as $\delta\bf{v}=0$ and $\delta\rho=\theta
mxy(\omega_x^2-\omega_y^2)/g$.  \cite{Lobo:05}. The subsequent
dynamical evolution can then be calculated using the time-dependent
polynomial ansatz
\begin{align}
  \delta\rho&=\frac{m}{g}[c(t)-\bf{r}\cdot\delta A(t)\bf{r}]\\
  \delta\bf{v}&=\delta B(t)\bf{r}
\end{align}
where $\delta A(t)$ is a symmetric and $\delta B(t)$ is a general
$2\times 2$ matrix and the chemical potential enters via the evolution
of the number $c(t)$ \cite{Lobo:05}. The numerical solution of these
equations shows an oscillation pattern which again is dominated by two
main frequencies, $\omega_{h1}$ and $\omega_{h2}$, and which we have
tabulated in Table \ref{tab:comparison}.

\begin{figure}[tb]
  \includegraphics[width=\linewidth,clip=true]{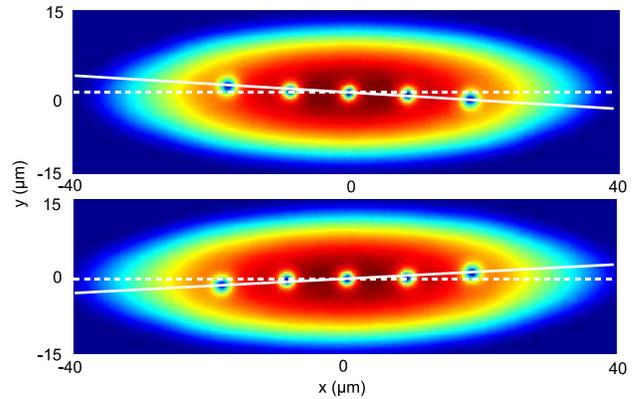}
  \caption{(Colour online) Density of the cloud at $t = 0.0072s$
    [upper panel] and $t=0.0108s$ [lower panel]. The full white line
    indicates the tilted axis of the vortex crystal and the broken
    white line the horizontal starting position. The oscillation of
    amplitude of the background cloud is too small to be visible in
    this plot.}
\label{fig:oscillations}
\end{figure}

While these are not, and were not expected to be, a perfect match for
the lattice oscillation frequencies, they are closer in value and in
particular show the same trend of increase and decrease. This shows
that the oscillations of the lattice excite and drive oscillations
within the background cloud, which in turn feeds back onto the lattice
dynamics to create the continuous and periodic behaviour. This is also
demonstrated in the fact that that the amplitude of the lattice
oscillations is significantly larger than the amplitudes of the
cloud. In Fig.~\ref{fig:oscillations} we show that the oscillation
angle of the lattice cloud is clearly much larger than the one for the
background cloud, which in fact stays very close to its initial
distribution. Nevertheless, the whole dynamics is highly predictable
and in particular the order of the vortices is maintained throughout.

\section{Dynamics in the Absence of Rotation}

It is interesting to investigate the robustness of the order of the
vortex lattice under even stronger perturbations of the background
cloud and we will therefore in the following investigate the dynamics
of the cloud in the absence of external rotation. From the results
above it should be clear that the background flow will be strongly
disturbed by the now non-rotating trapping potential and one might
expect the linear vortex crystal to break up into a random pattern. As
the system is finite, it is clear, though, that after some evolution the linear
form has to be re-assumed.

\begin{figure}[tb]
  \includegraphics[width=\linewidth,clip=true]{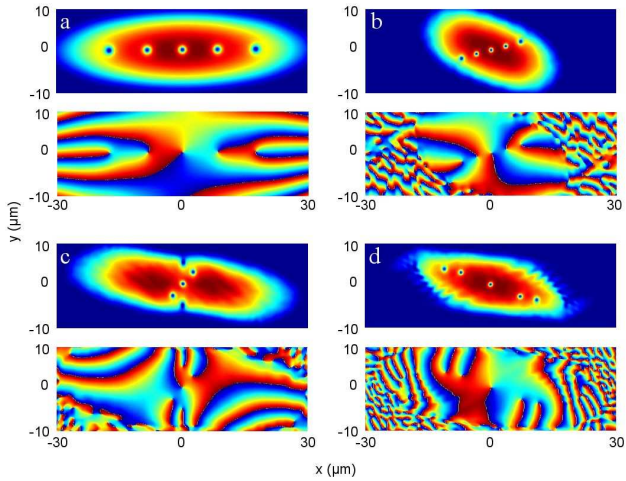}
  \caption{(Colour online) Dynamical evolution of the density and
    phase for a trap of aspect ration $\lambda=2.5$ for which the
    rotation has been switched off. Panel (a) shows the initial linear
    crystal at $t=0$ s and the subsequent panels show the times
    $t=0.03$ s (b), $t= 0.06$ s (c) and $t= 0.1$ s (d). Panel (c)
    shows the disturbed crystal as outer vortices pass the narrow
    direction of the condensate as they rotate around the central
    vortex and panel (d) shows the reformation of the linear
    crystal. At this time the left and right pairs of vortices have
    rotated around the central vortex, and the inner and outer
    vortices on each side have swapped position.  The phase plots show
    that the charge of the vortices is maintained throughout the time
    evolution.}
\label{fig:NonRot}
\end{figure}

\begin{figure}[tb]
  \includegraphics[width=\linewidth,clip=true]{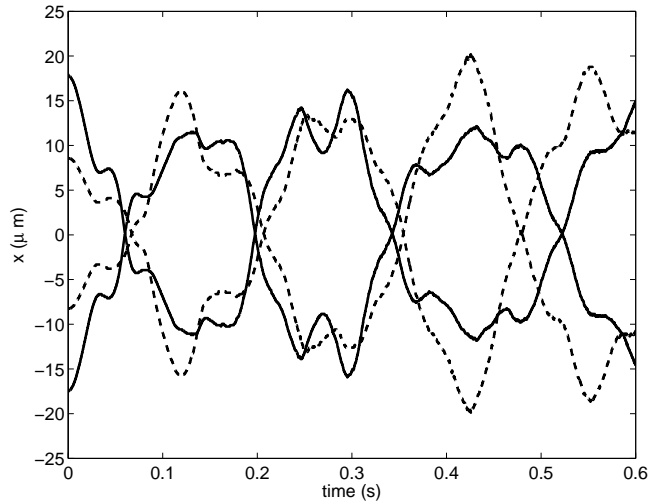}
  \caption{Position of the vortices along the $x$-direction for
    $\Omega = 0$ and for $\lambda = 2.5$.  The values for the
    initially outer pair of vortices are given by the full line and
    the inner pair of vortices by the broken. It can be seen that the
    paths remain symmetrical for the respective pairs, however they
    switch positions periodically.}
\label{fig:l_2_5_norot}
\end{figure}

In Fig.~\ref{fig:NonRot} we show an example of the time evolution of a
cloud with initially five vortices aligned along the soft axis (a) and
an aspect ratio of $\lambda=2.5$.  The vortex crystal can be seen to
rotate against the cloud, still maintaining its linear shape after
$t=0.03$ s (b). After a further $0.03$ s the linear geometry is
slightly disturbed when the vortices are passing through the narrow
direction of the condensate (c), but they finally assume the linear
shape again after about $0.1$ s (d). As for the case of maintained
trap rotation, the outer most pair of vortices follows a symmetrical
path, as do the inner pair of vortices. However once the narrow
section is passed the pairs on each side have changed their order for
the first time(see Fig.~\ref{fig:l_2_5_norot}) and this process will
be repeated over and over again during the ensuing evolution. The
vortices also regularly form a linear crystal again (see
Fig.~\ref{fig:l_2_5_norot}(d)), however not necessarily along the
horizontal axis. This behaviour persists for higher ratios of the
anisotropy, until the condensate itself decays too fast, due to the
anisotropic barrier it runs into, to be able to define a clear vortex
crystal.

\section{ Stability of vortices}

Let us finally briefly comment on the stability of vortices in the
lattice. As the systems we investigate are not rotationally symmetric,
they do not need to conserve angular momentum and it is possible for
the vortices to reverse their direction of flow
\cite{Ripoll:01,Watanabe:07}. However, this tunneling between
different rotational states is suppressed at larger interaction
strengths and as all our simulations are carried out in the
Thomas-Fermi limit, we have checked that all vortices keep their
initial rotational direction throughout the simulations. To demonstrate
this for one example, we show the phase distributions for the case of
the non-rotating cloud in Fig.~\ref{fig:l_2_5_norot}.

\section{Conclusion}

We have numerically investigated the dynamics of a linear vortex lattice
in an anisotropic, two-dimensional Bose-Einstein
condensate. Starting from the ground state of the Gross-Pitaevskii
equation, we have found that the vortices retain an ordered structure
both in the presence and absence of rotation for time-scales on the
order of 1 s. If the rotation of the cloud is maintained, the vortex
lattice remains linear and only carries out well defined oscillations
within the background cloud. With increased anisotropy oscillation
amplitude decreases until finally the crystal does not move within the
cloud. This allows for the use of the vortex lattices in applications
such as, for example, in quantum information. In the absence of
rotation we found that the outer vortices circle around the central
vortex, periodically reversing their order and the lattice and linear
crystals are periodically reformed for not too large
anisotropies. Once these get too strong, the background cloud gets too
distorted to still talk about a well defined vortex lattice.

\begin{acknowledgments}
  The work was supported by Science Foundation Ireland under project
  number 05/IN/I852. 
\end{acknowledgments}

% --------------------------------------------------------

\end{document}